\documentclass[lettersize,journal]{IEEEtran}
\usepackage{amsmath,amsfonts}
\usepackage{algorithmic}
\usepackage{algorithm}
\usepackage{array}
\usepackage[caption=false,font=normalsize,labelfont=sf,textfont=sf]{subfig}
\usepackage{textcomp}
\usepackage{stfloats}
\usepackage{url}
\usepackage{verbatim}
\usepackage{graphicx}
\usepackage{cite}
\usepackage{url}
\usepackage{multirow}
\usepackage{xcolor}
\usepackage{svg}
\usepackage{threeparttable}
\renewenvironment{IEEEbiography}[1]
  {\IEEEbiographynophoto{#1}}
  {\endIEEEbiographynophoto}

\usepackage{xspace}

\begin{document}
\title{Toward Communication-Efficient Space Data Centers: Bottlenecks, Architectures, and New Paradigms}

\author{Minghao~Sun,
        Zehui~Chen,
        Jinbo~Hou,
        Kezhi~Wang,~\IEEEmembership{Senior Member, IEEE,}
        Xiaoli~Chu,~\IEEEmembership{Senior Member, IEEE,}
        }


\maketitle

\begin{abstract}
The rapid growth of foundation model training and large-scale AI services has driven ground data centers toward unprecedented power densities, intensifying challenges in energy supply, cooling, and spatial scalability. Space Data Centers (SDCs) have emerged as a promising paradigm for hosting energy-intensive computing infrastructures in orbit, leveraging continuous solar energy and radiative cooling advantages. However, unlike ground facilities primarily constrained by power and site availability, SDCs are fundamentally limited by communication capability. The gap between petabit-scale internal data exchange in ground data centers and the gigabit-scale capacity of ground–space links forms a critical bottleneck.
This article systematically analyzes communication constraints in SDC architectures and explores semantic communication as a key enabling paradigm. By transmitting compact, task-relevant semantic representations instead of raw data, uplink pressure can be substantially reduced. 
The feasibility of communication-efficient orbital AI infrastructures is demonstrated through the evaluation of a multi-layer heterogeneous SDC framework consisting of relay satellites and orbital computing nodes operating under coupled energy and thermal constraints. The article further outlines open research challenges toward scalable deployment.


\end{abstract} 


\section{Introduction}

\IEEEPARstart{T}{he} rapid advancement of Artificial Intelligence (AI) has led to an unprecedented growth in data volume required for model training and continuous adaptation \cite{AI_SDC}.
At present, large hyper-scale data centers already operate at power levels close to 100 MW, and several proposals envision further scaling toward gigawatt-level computing facilities \cite{Starcloud_wp}.
Such power densities are comparable to those of medium-sized power plants, making grid integration and peak-load management increasingly challenging.
In this context, Space Data Centers (SDCs) leveraging satellite platforms have emerged as a promising paradigm for hosting large-scale computing resources outside densely populated regions, offering potential advantages in energy acquisition, cooling mechanisms, and spatial scalability \cite{SDC_survey}.
Recent industrial developments further illustrate this trend. In May 2025, Zhijiang Laboratory in China launched the first batch of computing satellites under the Three-Body Computing Constellation program. In November 2025, the Tongyi Qianwen (Qwen3) Large Language Model (LLM) was uploaded to the above satellites, enabling on-orbit deployment and inference.
Also in November 2025, StarCloud (formerly Lumen Orbit) launched a prototype orbital computing facility equipped with an H100 GPU, demonstrating onboard training with nano-GPT and inference testing using Google’s Gemma model. Google has announced Project Suncatcher, targeting the launch and hardware validation of prototype satellites to establish a space-based data infrastructure platform by 2027. Meanwhile, SpaceX has outlined plans to deploy computing-enabled satellites integrated with the Starlink network, forming a distributed architecture that dynamically coordinates communication and computational resources in orbit.
Such developments position SDCs as a new frontier in the evolution of AI infrastructure. However, this emerging computing paradigm introduces new challenges for communication architectures and system design, particularly under the coupled constraints of energy availability and thermal dissipation in space environments.
Existing studies on SDCs have largely examined energy harvesting, thermal management, or communication mechanisms in isolation. 
Yet systematic investigations that jointly incorporate these constraints into communication-oriented architectural design remain limited in SDC scenarios. This gap raises a fundamental question: Can current communication technologies support the massive heterogeneous data transmission of SDCs under multi-dimensional resource constraints?

In this article, we address this question through a systematic analysis of communication bottlenecks and representative service scenarios in SDC systems. 
We first identify the fundamental bottlenecks in existing SDC communication architectures and characterize the key service scenarios that determine their operational requirements. Based on this analysis, we identify candidate communication technologies capable of supporting SDC deployment across diverse application contexts.
A further contribution of this work is the incorporation of semantic communication as a foundational paradigm for SDC systems.
Instead of transmitting full-volume raw datasets, semantic communication conveys compact, task-relevant representations extracted from large-scale data. This shift reduces dependence on raw-data backhaul and alleviates pressure on ground-space link capacity.
To examine the feasibility of this approach, we develop a multi-layer heterogeneous satellite network architecture that integrates SDC nodes with communication satellites acting as relays between SDC nodes and ground stations. By incorporating semantic communication mechanisms, we evaluate the system's ability to sustain large-scale data transmission services under joint energy and thermal constraints.
The results offer practical insights into the design of communication-efficient SDC infrastructures.
Finally, we outline a research roadmap for semantic-enabled SDC frameworks, identifying key development stages and future research directions.

\section{Toward Sustainable and Efficient Computing in Space}
\begin{figure*}[t]
    \centering
    \includegraphics[width=2\columnwidth]{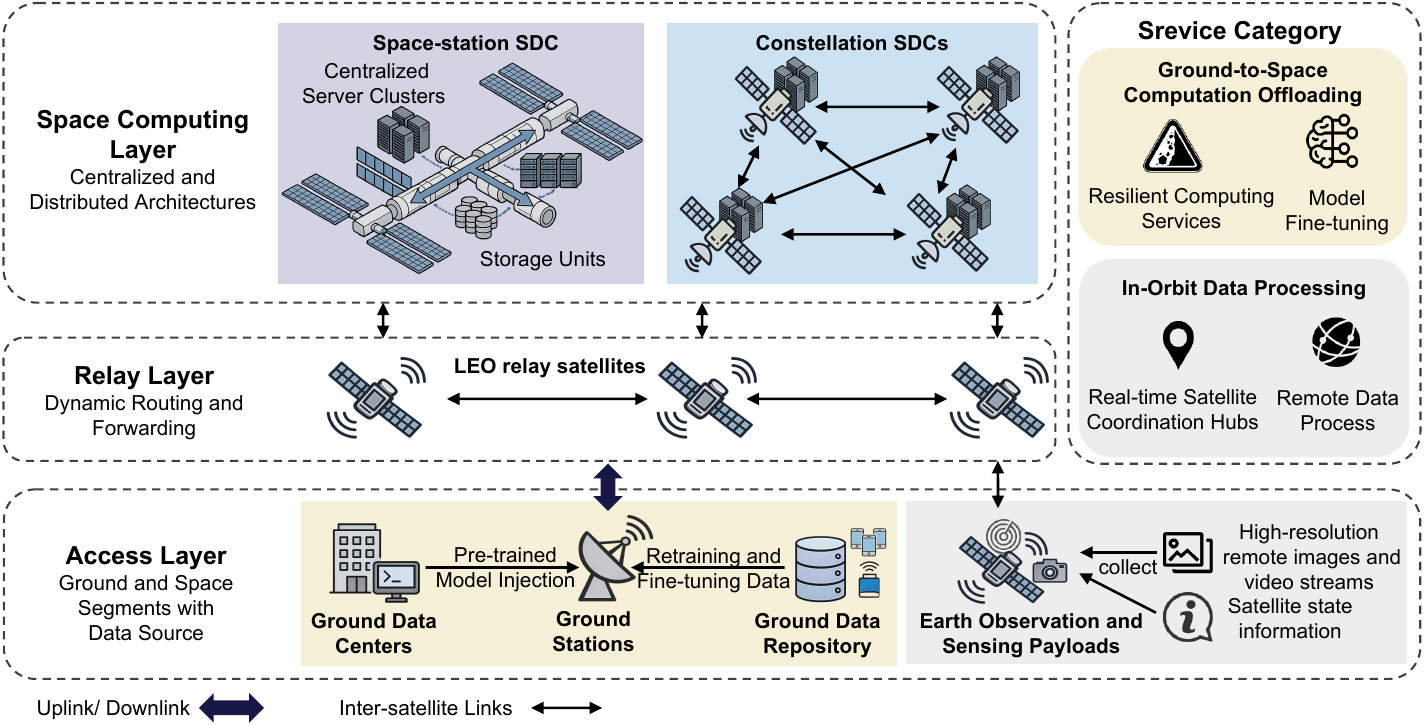}
    \caption{Space data centers: multi-layer architecture, and heterogeneous service types.}
    \label{intro}
\end{figure*}

In this section, we outline the system-level advantages of SDCs over conventional ground-based data centers and present a representative architectural and application vision, as illustrated in Fig.\ref{intro}.

\subsection{Why Space Data Centers?}
\textbf{Energy:}
At present, a key bottleneck of ground data centers lies in the growing gap between increasing data processing demand and the limited power supply of regional power grids.
In recent years, ultra-large-scale orbital solar arrays have been proposed to support space computing platforms, with the objective of providing long-term and reliable energy supplies for energy-intensive computing workloads.
The space environment offers nearly uninterrupted access to solar energy for large-scale computing infrastructures. In contrast to ground-based solar systems, which are constrained by day–night cycles, weather conditions, and geographical locations, solar radiation in orbital space is more stable and predictable.
By deploying large-scale solar arrays, SDCs can operate continuously and achieve more stable and efficient power utilization than ground photovoltaic systems \cite{solar_harvest}. 
Such sustained energy availability not only supports long-duration computing tasks but also enables more communication-intensive system designs. In particular, it creates the possibility of powering large-scale antenna arrays, high-frequency transceivers, and other energy-demanding communication modules, thereby facilitating higher-capacity space–ground and inter-satellite communications.
As a result, SDCs exhibit potential advantages in mitigating operational carbon emissions and environmental burdens. 
From a system-level perspective, the reliance of SDCs on solar energy during operation can substantially reduce dependence on fossil-fuel-based power sources. Compared to ground data centers that draw large amounts of electricity from regional power grids, SDCs may help alleviate energy structure pressures and carbon emission concentration issues associated with the centralized deployment of ultra-large-scale computing facilities, thereby offering a possible pathway toward more sustainable computing infrastructures.

\textbf{Cooling:}
Approximately 58\% of the electricity consumed by ground data center facilities is currently used for cooling systems rather than functional computing \cite{energy_report}. The deep-space radiative environment provides fundamentally different conditions for reducing cooling complexity and resource consumption at the system level.
Although the lack of physical media in space makes traditional heat dissipation methods relying on convection and water circulation inapplicable, its extremely low background temperature enables the heat generated by onboard computing systems to be dissipated directly into deep space through passive radiative heat transfer, reducing a significant amount of cooling water consumption. 
Moreover, the heat dissipation capacity of the facilities can be enhanced as the effective heat sink area expands, resulting in a relatively high scalability in heat dissipation capacity.

\textbf{Space Scalability:}
The vastness of orbital space allows computing infrastructures unprecedented flexibility in spatial scalability and deployment locations. 
Unlike ground-based data centers that are constrained by land availability, urban planning, and regional energy policies, SDCs can expand on demand through modular deployment and multi-satellite collaboration. 
Such an approach enables distributed space computing networks capable of supporting cross-regional and global computing services. This multi-satellite cooperative computing paradigm opens new possibilities for the elastic expansion and coordinated operation of future ultra-large-scale computing systems.

\subsection{Potential Architectures of SDCs}

\textbf{LEO Satellite Constellation Architecture:} 
Within the constellation, satellites assume complementary roles to form a multi-layer distributed computing architecture in orbit. 
Computing satellites equipped with onboard servers provide primary processing capability, while dedicated communication satellites extend ground coverage, support task offloading, and act as relay nodes between computing satellites and ground stations. 
Dynamic interconnections among the computing nodes enhance architectural flexibility, enabling new computing satellites to be added and aging platforms to be replaced without interrupting service continuity.

\textbf{Space Station–Hosted Data Center Architecture:}
This approach deploys server clusters directly on a space station, effectively establishing a centralized data center in orbit. 
A key advantage is that its structural design and maintenance requirements closely resemble those of conventional ground-based data centers, thereby reducing the need for additional maintenance mechanisms.
Moreover, co-locating computing resources on a single space platform centralizes processing capability and lowers the overhead associated with multi-satellite trajectory coordination and inter-satellite communication, simplifying overall system management.

\subsection{What Can SDCs Do?}

\textbf{In-Orbit Data Processing:}
Such services leverage onboard pre-processing and intelligent analysis of remote sensing images or video streams to reduce downlink data volume and improve real-time responsiveness. 
In addition, SDCs can act as coordination hubs for large-scale space-based systems, using real-time satellite state information to optimize orbital trajectories and service allocation. This enables dynamic, near real-time coordination across satellite clusters. 
In this scenario, latency-sensitive tasks are prevalent, making efficient multi-satellite routing and communication resource allocation critical.

\textbf{Ground-to-Space Computation Offloading:}
Through ground–satellite links, data can be offloaded for in-orbit processing, where continuous solar energy supports sustained large-scale computation. This architecture is particularly suitable for latency-tolerant yet computation-intensive workloads, such as large-model training, retraining, and multi-task fine-tuning.
Beyond dedicated space-based services, SDCs can also serve as complementary computing resources to ground data centers. For instance, during peak demand periods or in the event of natural disasters, selected workloads can be shifted to orbital platforms to relieve pressure on ground infrastructure and enhance overall system resilience.

\section{Bottlenecks and Solutions}

In this section, we examine the fundamental bottlenecks limiting current SDC deployments and outline potential technical solutions along with their applicable scenarios.

\subsection{Challenges Facing SDCs}
\textbf{Communication Capacity:}
The rapid rise of LLMs has made training and inference for foundation models core workloads in modern data centers. Supporting model pre-training, continual adaptation, and large-scale inference requires massive datasets and extensive computational resources, driving rapid expansion in data center capacity and scale.
In ground data centers, infrastructure expansion is primarily constrained by power supply and site availability. In contrast, SDCs face system-level limitations that are more closely tied to communication capability. 
According to the 2024 Google Cloud report \cite{google_2024}, the data throughput of a single aggregation block in modern data center networks has reached approximately 200 Tb/s. When dozens of such aggregation blocks are interconnected, the aggregate data exchange capacity can scale to the Pb/s level.
However, due to limited transmit power, constrained spectrum resources, severe propagation losses over long distances, and intermittent connectivity, existing ground–satellite access links typically provide capacities of only tens to hundreds of Gb/s \cite{capacity}.
In Table \ref{single_uplink_capacity}, we summarize the uplink capacity supported by several representative satellite platforms and communication technologies currently under development or deployment. The comparison shows a substantial gap between the capabilities of current communication systems and the data transmission demands envisioned for SDC operation.
This gap makes it difficult for SDCs to support computing and data-processing capabilities comparable to those of ground data centers.
Consequently, communication capacity represents a key bottleneck for the practical deployment and large-scale operation of SDCs.

\textbf{Software-Defined System Scheduling:}
Software-defined scheduling faces multidimensional challenges in SDC environments. Due to the time-varying and intermittent connectivity of satellite links, task feedback, control signals, and model updates cannot always be disseminated across distributed nodes in real time. As a result, during no-contact intervals, different nodes may temporarily operate under inconsistent software versions or configuration states. Task data may accumulate in local buffers and be transmitted in batches once connectivity is restored, leading to bursty traffic patterns and transient congestion. 
To ensure reliable system operation, scheduling mechanisms must therefore accommodate not only dynamically varying channel capacity but also delayed state synchronization. This requires elastic buffering strategies and priority-aware scheduling policies to maintain stable task execution under asynchronous updates and propagation delays, while preventing adverse impacts on task-related transmission rates.

\textbf{Data Security:}
The openness and broadcast nature of SDCs' long-distance ground-to-satellite wireless transmissions increase the risk of data leakage and eavesdropping \cite{data_security}.
Future SDC infrastructures may support multiple organizations, services, and heterogeneous tasks operating on shared orbital resources. 
In such multi-tenant, diverse-service environments, security mechanisms designed to protect a specific task may introduce additional overhead that degrades the performance of other workloads. Moreover, ensuring data privacy and establishing trust across different stakeholders adds further complexity to system design.

\begin{table*}[t]
\caption{Single-uplink capacities under different platforms and technologies}
\label{single_uplink_capacity}
\centering
\footnotesize
\setlength{\tabcolsep}{4pt}
\renewcommand{\arraystretch}{1.15}
\begin{tabular}{|c|c|c|}
\hline
\textbf{Platform / technology} & \textbf{Single uplink capacity} & \textbf{Source} \\
\hline
Starlink (typical service) & 5--20 Mbps &
Starlink, Specifications\\
\hline
Starlink lower-speed tier & 20 Mbps &
Starlink, Network Update \\
\hline
SES O3b mPOWER (per terminal) & Multi-Gbps &
SES, O3b mPOWER \\
\hline
SES O3b mPOWER IP Transit & Up to 10 Gbps &
SES, O3b mPOWER IP Transit Executive Guide\\
\hline
Airbus / CNES TELEO optical feeder link & 10 Gbps &
Airbus, TELEO Announcement\\
\hline
NASA next-generation optical uplink & $>$100 Gbps &
NASA, Next-Generation Optical Communications Relay \\
\hline
NASA optical relay pathfinder & 100 Gbps &
NASA, Optical Relay Pathfinder \\
\hline
\end{tabular}
\end{table*}

\subsection{SDC Enabling Technologies}
Within the conventional communication paradigm, the most critical structural bottleneck in SDC systems is limited communication capacity, which fundamentally restricts the feasibility of ground-to-space task delivery at scale. Many other challenges only become pronounced once this primary bottleneck is alleviated and the system begins to support simultaneous transmission and coordination of numerous heterogeneous tasks. Accordingly, this subsection focuses on the key technologies that may help overcome the communication-capacity limitation in current SDC architectures.

\textbf{Semantic Communication:}
Semantic communication has emerged as a promising approach to addressing the communication capacity bottleneck in SDC scenarios. Unlike conventional communication systems that emphasize bit-level accuracy, semantic communication focuses on meaning fidelity or task utility. 
The objective is not to perfectly reconstruct every transmitted bit, but to ensure that the received information preserves its semantic intent or task relevance.
By proactively filtering redundant or task-irrelevant content under bandwidth constraints, semantic communication can significantly improve effective throughput and communication efficiency \cite{semantic_1}.
A typical semantic communication system includes a shared knowledge base or semantic model, with the receiver using this shared knowledge for semantic inference and information recovery.
In this regard, SDCs offer a unique advantage: their substantial computational power and large-scale storage capacity allow maintenance and continuous updating of massive semantic models and knowledge repositories. With these resources, a semantic receiver can actively retrieve factual or structured knowledge from SDC-hosted knowledge bases to assist in semantic decoding and reasoning \cite{semantic_2}.
In SDC-oriented AI workloads, in-orbit fine-tuning of deployed LLMs represents a major task. 
Such a training objective is typically well defined and does not require preserving the complete data distribution. Samples with marginal contribution to model updates can be discarded, while only high-impact gradient information or task-relevant features are transmitted \cite{fine_tune}.
This aligns naturally with task-oriented semantic communication, which prioritizes task performance over bit-level reliability. 
Instead of minimizing bit error rate, the communication process is designed to maximize task success probability or minimize task loss. By transmitting high-value samples or semantically compressed feature representations, link utilization can be significantly reduced.
Semantic encoders extract task-relevant features and jointly optimize task success rate, training loss, and downstream performance metrics. Such techniques can reduce communication overhead by up to 60\% compared with raw data transmission \cite{semantic_overhead}.

\textbf{Multi-Antenna Beamforming for SDC Networks:}
Multiple-Input Multiple-Output (MIMO) is a communication technology that employs multiple antennas at the transmitter and receiver to improve link performance.
By leveraging array gain, beamforming, or spatial multiplexing, MIMO can increase the effective capacity per link to the Gb/s level, particularly in scenarios with a high signal-to-noise ratio \cite{mimo}.
In SDC systems, stable energy supply and strong onboard processing capabilities make it feasible to support large antenna arrays, high-frequency transceivers, and real-time beam control. Meanwhile, the predominantly line-of-sight characteristic of space-ground links favors highly directional transmission, allowing beamforming to improve link budget and communication capacity more effectively. This makes multi-antenna beamforming a promising solution for relieving communication bottlenecks in SDC networks.

\textbf{Inter-Satellite Laser Communication:}
Inter-Satellite Laser (ISL)  communication is a high-capacity optical transmission technology that employs narrow laser beams to establish direct space-to-space links between satellites. According to the 2025 \textit{IEEE Spectrum} report \cite{laser}, ISL links can achieve data rates of up to 400 Gbps.
Compared with conventional radio-frequency links, ISL links offer significantly higher bandwidth and lower transmission latency. Such high-throughput connectivity provides a critical networking foundation for distributed computing and real-time coordination within SDC architectures, enabling satellites to operate as an integrated orbital computing system.

\section{A Communication-Efficient SDC Framework Incorporating Semantic Communication Under Coupled Energy and Thermal Constraints}
In practice, SDC operation requires co-design across three tightly coupled domains: energy, thermal management, and communication.
Computing workloads consume power and generate heat, while thermal constraints limit sustainable compute intensity, and intermittent ground-space connectivity governs when data, model updates, and control signals can be exchanged.
Consequently, the system must be managed through both long-term window-level scheduling and real-time computing control mechanisms.
Accordingly, in this section, we first introduce the fundamental network architecture of SDC systems. 
Based on this framework, we then compare semantic communication with conventional bit-level communication in terms of uplink utilization and ground station (GS) energy consumption under the joint constraints of power supply, computational demand, and thermal dissipation.

\begin{figure*}[t]
    \centering
    \includegraphics[width=2\columnwidth]{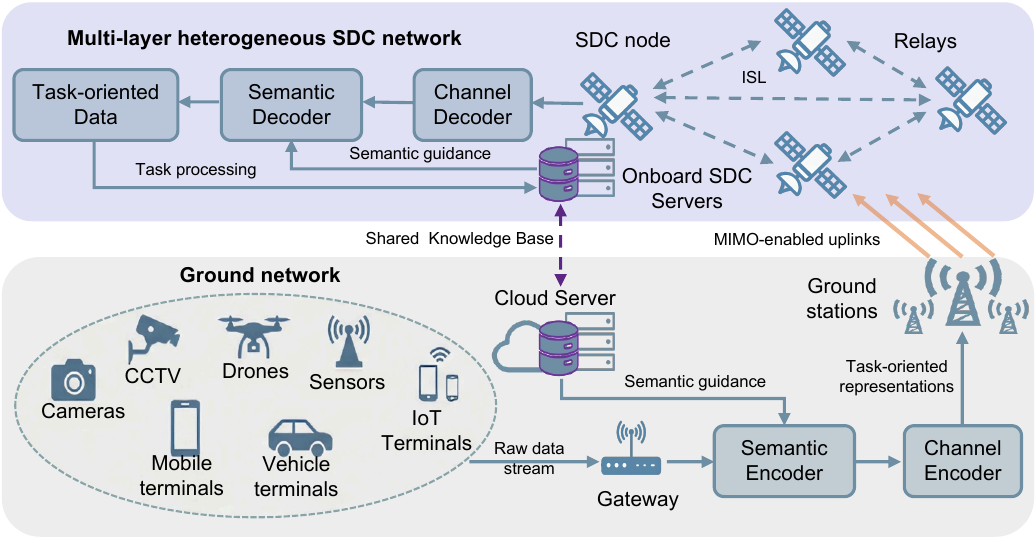}
    \caption{Architecture of a Semantic Communication-enabled Multi-Layer Heterogeneous SDC Communication Framework.}
    \label{system}
\end{figure*}
\subsection{A Multi-layer Heterogeneous SDC Framework} 
We propose a multi-layer heterogeneous network architecture that consists of a space computing layer, a relay layer, and an access layer, as shown in Fig. \ref{intro}.
Within this framework, GSs communicate directly with SDC platforms when coverage is available; otherwise, outside the SDC service footprint, ground traffic is first offloaded to relay satellites and then forwarded to the SDC.
Multi-satellite scheduling mechanisms enable dynamic workload allocation across computing nodes, allowing tasks to be routed, migrated, or replicated based on resource availability and service requirements. 
This flexibility enables the system to accommodate time-varying demand and heterogeneous workloads.
As illustrated in Fig. \ref{system}, a semantic communication (SemCom)-enabled SDC framework is designed to support large-volume ground-space transmission, where ground-space links adopt SemCom and MIMO to enhance effective link capacity and data throughput, while ISL links provide high-bandwidth, low-latency connectivity across the constellation, forming a backbone for distributed in-orbit computation.
Furthermore, the dynamic interconnection among satellites enhances architectural flexibility, allowing new computing nodes to be incrementally integrated and aging platforms to be replaced without disrupting ongoing services. This modular and evolvable structure supports scalable expansion and long-term system upgrades.

\subsection{Case Study for Ground-Space Computation Offloading}

To evaluate whether the proposed framework can sustain high-throughput SDC operation, we compare conventional bit-level communication (BitCom) with SemCom within a multi-layer multi-LEO satellite network. Specifically, we examine the ground-space uplink capacity required by each scheme to achieve a throughput matching the maximum sustainable data processing volume.

In the simulated framework, the total SDC service duration is set to 10s, divided into 10 time slots of 1s each. The ground segment consists of 30 GSs, while the space segment includes 24 communication satellites and 1 SDC operating at an altitude of 500 km.
The communication satellites are assumed to have no onboard computing capability; instead, they forward all received semantic information to the SDC, which is equipped with multiple H100 GPUs, through 400 Gbps ISL links for processing. 
Each GS is assumed to access up to two orthogonal ground-satellite channels simultaneously, where each satellite can maintain up to six simultaneous inter-satellite links.
A load-aware scheduling mechanism is incorporated, such that when a relay satellite experiences high traffic load, data are preferentially routed to alternative satellites with lower load whenever connectivity permits.

The SDC platform is equipped with a square solar array and a rear-mounted radiative panel for heat dissipation to minimize structural footprint. Let the total solar power generation be denoted by $P$, and the computing power consumption by $E_1$. Since most of the electrical power consumed by GPUs is converted into heat, the thermal generation rate is also approximated as $E_1$. The processing energy consumption is assumed to be proportional to the processed data volume $X$ \cite{energy}.
Since the energy consumption associated with signal reception is significantly smaller than the computing power demand under the considered system configuration, it is neglected in the subsequent analysis for simplicity.
The radiative panel provides a net heat dissipation capacity of $E_2$ after excluding the/ absorbed solar radiation \cite{Starcloud_wp}. 
Due to the need for active coolant circulation, we assume that the pumping power consumption accounts for 2\% of the radiative dissipation power. 
Therefore, sustainable system operation requires satisfying the following conditions: $E_1 + 0.02E_2 \leq P$ and $E_1\leq E_2$.
The first constraint ensures sufficient solar power for both computation and cooling, while the second guarantees that generated heat does not exceed the radiative dissipation capacity.
Under a fixed power budget $P$, these constraints determine the maximum sustainable data processing volume $X_{\rm{max}}$. 

In the semantic communication configuration, each GS employs a trainable ResNet18-based encoder to transform each CIFAR-10 image into a compact semantic packet. Specifically, the model adopts a modified ResNet18 backbone for feature extraction, using a $3×3$ input convolution and removing the initial max-pooling layer. The extracted features are then compressed through a linear bottleneck into an 80-dimensional latent vector, quantized to 3 bits per dimension via a straight-through estimator, and combined with 16 bits of fixed overhead to form a 256-bit semantic representation.

\begin{figure}[t]
    \centering
    \includegraphics[width=3.4in]{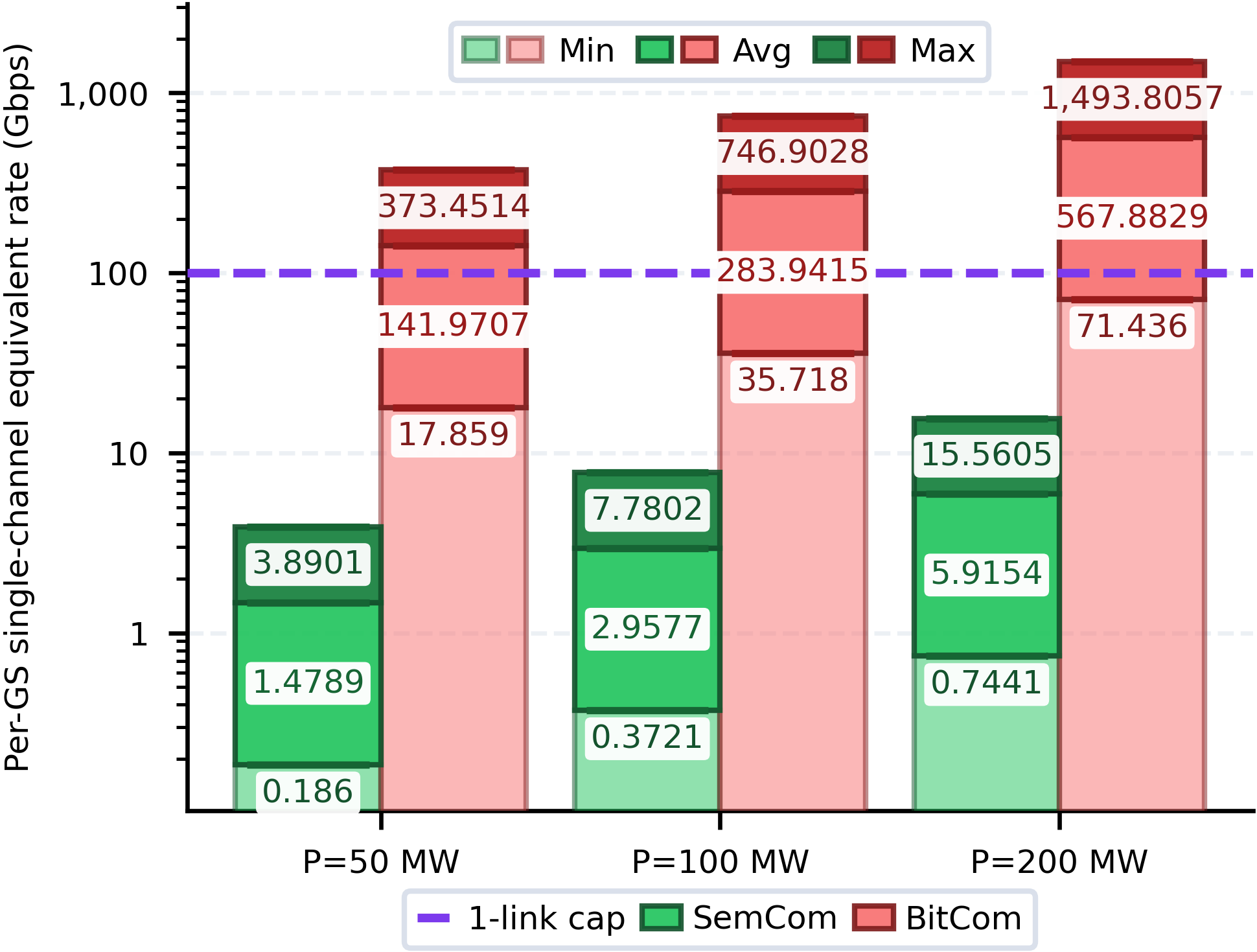}
    \caption{Per-GS equivalent single-channel data rate required to achieve $X_{\rm{max}}$ under different SDC power budgets.}
    \label{result1}
\end{figure}
\begin{figure}[t]
    \centering
    \includegraphics[width=3.4in]{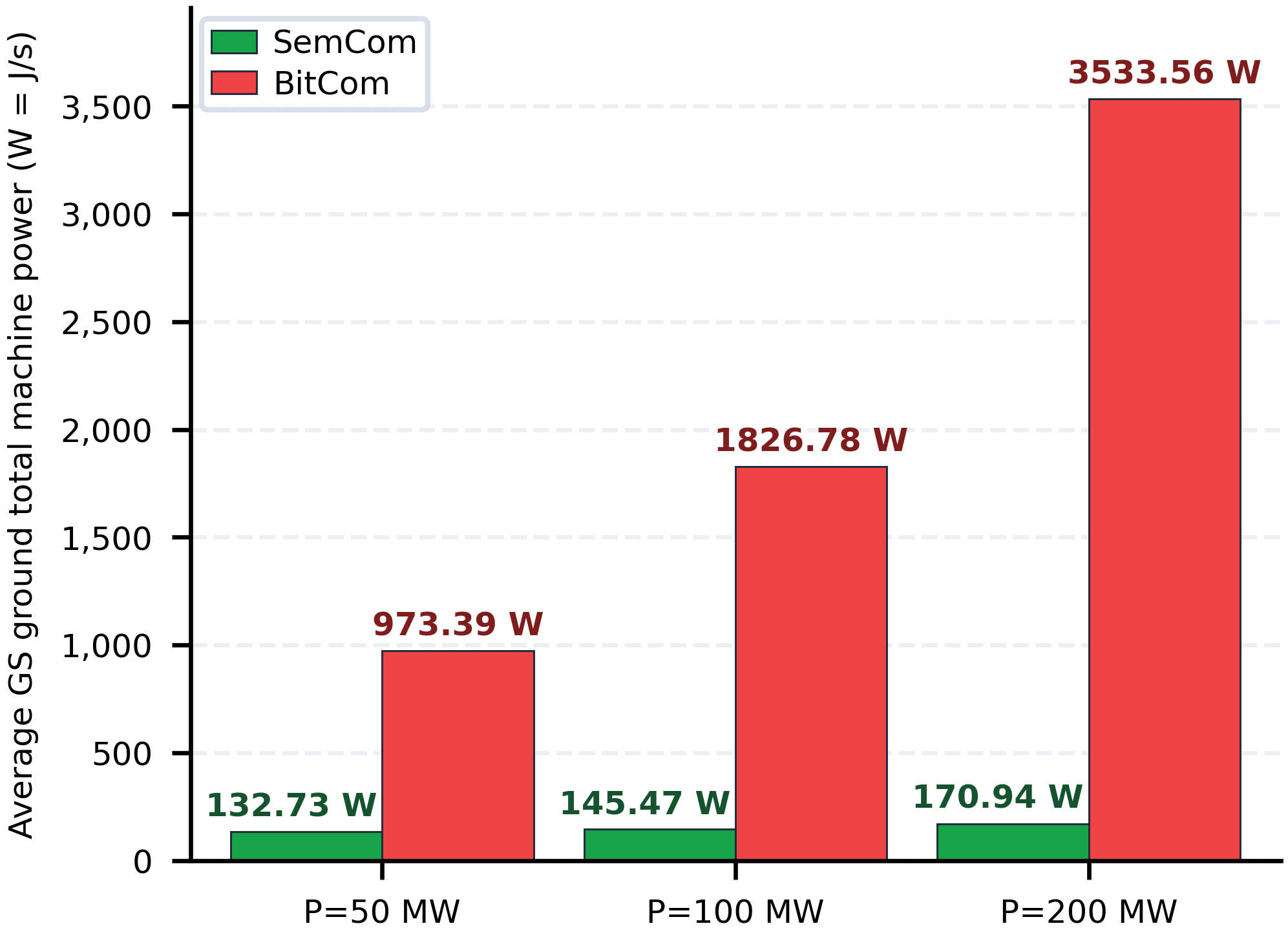}
    \caption{Average power consumption of GSs to achieve $X_{\rm{max}}$ under different SDC power budgets.}
    \label{result2}
\end{figure}
 
Fig. \ref{result1} shows the equivalent single-channel data rate per GS required by SemCom and Bitcom to achieve $X_{\rm{max}}$ under different SDC power budgets.
According to the demonstrations reported in the literature, a single ground-space link can support data rates of up to 100 Gbps, which is indicated by the horizontal reference line in Fig.\ref{result1}.
The simulation results show that, to sustain an SDC with a processing capability exceeding 50 MW, conventional bit-level communication requires ground-space link capacities beyond the 100 Gbps limit. This indicates that, under current link constraints, bit-level transmission alone cannot support full-load SDC operation.
In contrast, when semantic communication is employed, the required link utilization is dramatically reduced by more than 98\% in the evaluated scenario, while maintaining a task accuracy of 94.43\%. 

Given that SDC platforms are powered by solar arrays capable of delivering megawatt-level energy, energy constraints are more likely to arise at GS rather than in orbit. To capture this effect, we evaluate the transmission energy consumption at GSs for delivering a data volume of $X_{\rm{max}}$.
Fig. \ref{result2} shows the average power consumption of GSs using SemCom or BitCom to achieve $X_{\rm{max}}$ under different SDC power budgets.
The simulation results demonstrate that, despite the additional computational overhead introduced by the semantic encoder, the total energy consumption of semantic communication remains substantially lower than that of conventional bit-level transmission.
These findings suggest that semantic communication can substantially alleviate ground-space link pressure while reducing energy cost, making it a promising solution for supporting high-throughput SDC deployments under realistic communication constraints.

\section{Future Research Directions}

Realising the full potential of semantic communication for supporting SDCs still presents several open challenges. Based on the multi-layer heterogeneous SDC framework evaluated in this study, we outline a set of future research directions, identify key challenges, and discuss potential solution pathways.

\textbf{Token-Level Semantic Reconstruction:}
Large-scale pre-training of LLMs requires learning general data distributions from fully tokenized sequences. In such settings, defining task importance or selectively filtering tokens without compromising distributional integrity becomes highly challenging. 
An alternative approach is semantic-oriented communication, which prioritizes reconstruction fidelity at the semantic level rather than strict task-specific objectives. While this strategy may alleviate the limitations of task-oriented filtering, it introduces new challenges, including distribution shift, over-normalization effects, hallucinated reconstruction errors, and reduced reproducibility across training runs.
If these challenges can be systematically addressed, semantic communication could potentially extend beyond fine-tuning scenarios to support a broader range of LLM training workloads, thereby substantially reducing ground–space link pressure in large-scale SDC deployments.

\textbf{Cross-Task Knowledge Generalization:}
Semantic communication relies on shared knowledge bases or semantic models for encoding and decoding. Existing semantic communication systems typically construct task-specific knowledge repositories tailored to predefined objectives, which limits their flexibility across heterogeneous workloads. Recent advances in LLMs offer new opportunities to address this limitation. Generative foundation models can facilitate knowledge transfer and dynamic knowledge adaptation across tasks. 
Such a generative-assisted knowledge framework may improve cross-task generalization and reduce the need for manual knowledge engineering, thereby enhancing the scalability and flexibility of semantic communication in SDC environments.

\textbf{Long-Term Security Risks:}
Unlike one-off transmissions, semantic information (SI) can be cached, aggregated, and reused for training or retrieval. Even without access to the original data, an adversary may exploit large volumes of accumulated SI as a queryable semantic index to infer sensitive attributes or critical patterns.
In SDC communication architectures, where semantic data may be stored, shared, and reused across multiple transmissions and tasks over extended periods, the security focus of semantic communication shifts from protecting transient communication channels and bitstreams to safeguarding SI as a long-term data asset, thereby requiring fundamentally new security mechanisms.

\section{Conclusion}
In this article, we have presented a systematic analysis of communication constraints in SDC architectures and examined representative service scenarios under joint energy and thermal considerations. By incorporating semantic communication as a foundational paradigm, we have demonstrated that task-oriented semantic representations can significantly alleviate uplink pressure while preserving task performance. Through a heterogeneous multi-layer architecture integrating relay satellites and inter-satellite laser links, we have further shown that communication-efficient SDC operation is feasible under realistic system constraints.

Despite these promising findings, several open challenges remain, including token-level semantic reconstruction for large-scale pre-training, adaptive cross-task knowledge generalization, and long-term security risks associated with accumulated semantic information. Addressing these issues will be essential for extending semantic communication beyond fine-tuning scenarios and enabling scalable orbital AI infrastructures. Overall, the practical viability of future gigawatt-scale SDC systems will depend not only on computational capability, but also on communication-efficient system design.

\section*{Acknowledgment}
This work was supported in part by the Horizon Europe Research and Innovation Program under grants 101086219 and 101086228, the UK EPSRC grants EP/X038971/1 and EP/Y028031/1, Innovate UK COMET project (No: 10099265), and Royal Society Industry Fellowship (IF$\setminus$R2$\setminus$23200104).
Minghao Sun and Zehui Chen are co-first authors.
Xiaoli Chu is the corresponding author.

\bibliography{ref}
\bibliographystyle{IEEEtran}  
 \vspace{-14 mm}
  \begin{IEEEbiography} {Minghao Sun} is pursuing the Ph.D. degree with the School of Electronic and Electrical Engineering of the University of Sheffield, U.K. (email:msun39@sheffield.ac.uk)
\end{IEEEbiography}
 \vspace{-14 mm}
  \begin{IEEEbiography} {Zehui Chen} is pursuing the Ph.D. degree with the School of Electronic and Electrical Engineering of the University of Sheffield, U.K. (email:zchen199@sheffield.ac.uk)
\end{IEEEbiography}
 \vspace{-14 mm}
\begin{IEEEbiography} {Jinbo Hou} is pursuing the Ph.D. degree with the School of Electronic and Electrical Engineering of the University of Sheffield, U.K. Jinbo is also with Brunel University of London, U.K. (email:jhou9@sheffield.ac.uk)
\end{IEEEbiography}
 \vspace{-14 mm}
\begin{IEEEbiography} {Kezhi Wang} is a Professor with the Department of Computer Science, Brunel University of London, U.K. (email:kezhi.wang@brunel.ac.uk)
\end{IEEEbiography}

 \vspace{-14 mm}
\begin{IEEEbiography} {Xiaoli Chu} is a Professor with the School of Electronic and Electrical Engineering, University of Sheffield, U.K. (email:x.chu@sheffield.ac.uk)
\end{IEEEbiography}

\end{document}